# Designing a Resource Broker for Heterogeneous Grids

Srikumar Venugopal, Krishna Nadiminti, Hussein Gibbins and Rajkumar Buyya

Grid Computing and Distributed Systems Laboratory
Dept. of Computer Science and Software Engineering
The University of Melbourne, Australia

*{srikumar, kna, hag, raj}@csse.unimelb.edu.au*

**Abstract:**
Grids provide uniform access to aggregations of heterogeneous resources and services such as computers, networks and storage owned by multiple organizations. However, such a dynamic environment poses many challenges for application composition and deployment. In this paper, we present the design of the Gridbus Grid resource broker that allows users to create applications and specify different objectives through different interfaces without having to deal with the complexity of Grid infrastructure. We present the unique requirements that motivated our design and discuss how these provide flexibility in extending the functionality of the broker to support different low-level middlewares and user interfaces. We evaluate the broker with different job profiles and Grid middleware and conclude with the lessons learnt from our development experience.

## 1 Introduction

Grids [1] enable collaboration via virtualisation and sharing of resources and processes, owned, managed and used by multiple organisations and individuals from across the globe. Originally born out of the needs of high-end scientific investigations, this model of computing has come to be used in a wide range of areas including scientific research, engineering, business, finance and manufacturing. Grids consist of distributed heterogeneous resources such as high-end supercomputers, clusters, storage repositories, databases and scientific instruments connected by high-speed networks thereby representing a wide variety of computing platforms and software systems.

The capabilities that need to be provided in order to realise a Grid environment include: uniform authentication and authorisation; resource management and job submission; large-scale data management and transfer; and resource allocation and scheduling. Software tools and services that provide these capabilities are collectively called *Grid middleware*, which mediate between users and the underlying Grid fabric consisting of heterogeneous computing and storage resources connected by networks of varying capabilities. Figure 1 shows the evolution of Grid application development using Grid middleware.

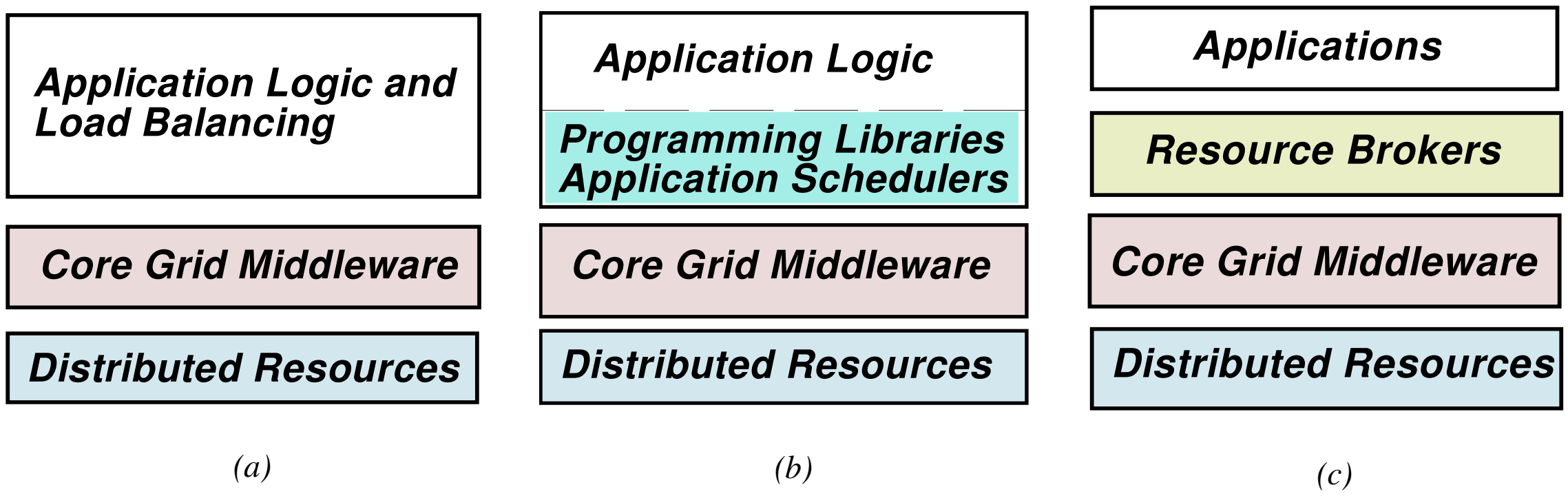

*Figure 1: Evolution of application development on Grids. (a) Direct to core middleware (1995-). (b) Through programming libraries (1999-). (c) Using user-level middleware (2000-).*

The first generation of Grid middleware aimed to present a secure and standard method of invocation that abstracted the underlying heterogeneity of distributed resources. These *core grid middleware* such as Globus [2] and Legion [3] (Figure 1(a)) provided services for performing low-level Grid functions such as data access, job submission and authorisation. Some of the early Grid applications directly invoked the functionalities presented by these *core middleware* through their APIs. However, they were still too complex and low-level to become popular for general application development. These were followed by programming libraries such as NetSolve[4], Ninf[5], Cactus[6] and GrADS [7] (Figure 1(b)) which provided a software environment to create applications that accessed distributed services in a transparent fashion. While these libraries did a good job of shielding the developer from having to deal with varying Grid conditions, significant effort was still required to develop schedulers and task managers for each application. Projects such as AppLeS (Application Level Schedulers) [8] produced some of the early work in this regard. The next step was to move the scheduling algorithms to a generic framework that would also provide capabilities such as resource selection, job monitoring and data access to any application. Such frameworks are called *resource brokers* (Figure 1(c)) and some examples of these are Nimrod/G[10], AppLeS Parameter Sweep Template (APST)[11], Condor-G[12], and the Gridbus broker which is the subject of this paper.

A previous publication [13] introduced the Gridbus broker as a scheduler for distributed data-intensive applications on global Grids. A case study using a data-intensive High Energy Physics application was also presented to illustrate the use of the broker. On the other hand, this paper takes a detailed look at the architecture and design of the broker and discusses how these were motivated by the requirements of working in a service-oriented, heterogeneous Grid environment where each user has his own Quality-of-Service expectations to be satisfied. The Gridbus broker is then compared with related projects to highlight its uniqueness and the differences in function that have resulted from its design. Finally, the paper presents measurements that show the impact of the design principles on the performance of the components of the broker in different scenarios.

## 1.1 Resource Brokers: Challenges and Requirements

The characteristics of Grids such as large number of services with multiple configurations, dynamic resource conditions and users with varying application requirements introduce unique challenges that are listed as follows:

**Service Heterogeneity:** With the introduction of the OGSA [14], Grids have progressed from being aggregations of heterogeneous resources to collections of stateful services. These services can be grouped into several categories such as job submission and monitoring, information, data management and application deployment. Rapid developments in this field have meant that middleware itself changes frequently, and therefore service interface changes are a norm rather than the exception. While these interfaces are being standardised in fora such as the Open Grid Forum (OGF), a lot of Grid applications are being developed and deployed in industry and academia using existing non-standardised interfaces. Therefore, due to this heterogeneity of service interfaces, supporting diverse service semantics is still a serious challenge.

**Variety of Application Models:** Grid applications tend to follow a variety of models such as bag of tasks, workflows, independent jobs and hosted application services. However, these are still required to interact with the same set of service interfaces. Enabling this interaction requires reconciliation between different application directives and constructs. Also, applications may invoke services in a variety of ways. A brokering system must avoid imposing constraints on applications as far as possible so as to not limit its own applicability.

**Multiple User Objectives:** Applications and users may wish to satisfy different objectives at the same time. Some possible objectives include receiving results in the minimum possible time or within a set deadline, reducing the amount of data transfer and duplication, or ensuring minimum expense for an execution or minimum usage of allocated quota of resources. Different tasks within an application may be associated with different objectives and different QoS (Quality of Service) requirements. A brokering system must, therefore, ensure that different scheduling strategies meeting different objectives can be employed whenever required.

**Interface Requirements:** The interface presented to the user may take several forms. Many scientists are comfortable with traditional command-line tools and require that Grid tools should be command line and script-friendly as well. Web portals that allow users to invoke Grid and application capabilities within one interface have gained popularity in recent times as they enable portability of working environments.

Recent applications also seamlessly access Grid functions whenever required by invoking Grid/Web services or Grid middleware APIs. A resource broker should be able to support as many of these interfaces as possible in order to be useful to the largest community possible.

**Infrastructural Concerns:** The quintessential properties of Grid environments such as absence of administrative control over resources, dynamic system availability and high probabilities of failure have been described extensively in previous publications [15][16]. A resource broker has to be able to handle these properties while abstracting them as much as possible from the end-user. This is a significant challenge to developing any Grid middleware.

The following sections present the architecture, design and implementation of a Grid resource broker that takes into account the challenges mentioned so far in order to abstract the vagaries of the environment from the end-user.

## 2 Architecture of the Gridbus Broker

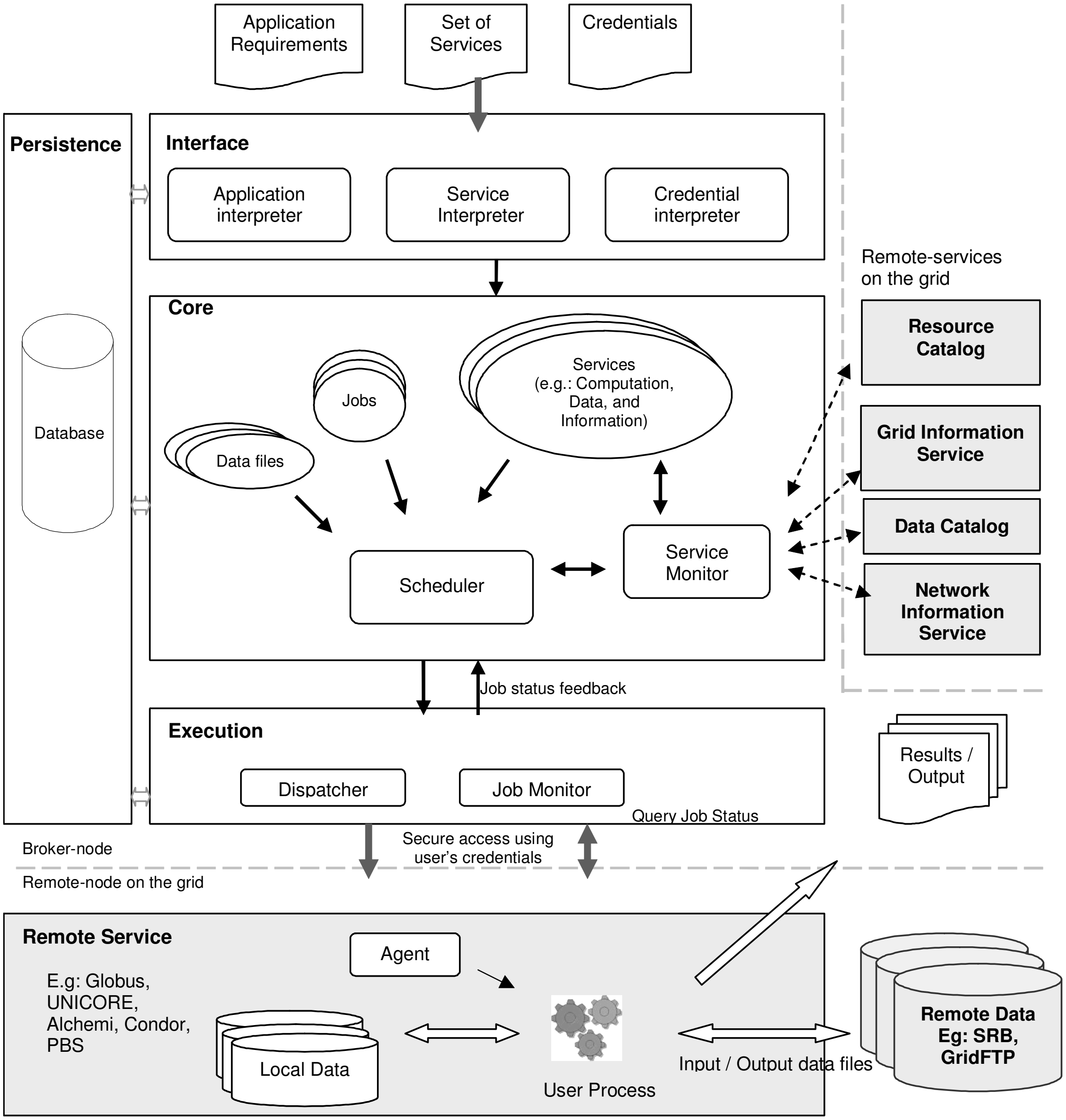

*Figure 2: Gridbus broker architecture and its interaction with other Grid entities.*

The architecture of the Gridbus broker is shown in Figure 2, and consists of three primary layers (Interface, Core and Execution) with the database as a persistent backend throughout. This architecture isolates the logic (at the Core) from interactions with both the user/application interfaces and remote Grid services. The components of the broker are grouped into the tiers based on the level of abstraction they provide from the underlying Grid resources. The overall flow of control is from top to bottom while any events and exceptions that occur during execution are filtered by each tier from the bottom to the top. The tiers are described in detail in the following paragraphs.

## 2.1 Interface Layer

Applications, web portals and other such interfaces external to the broker interact with the components of the Interface layer. The inputs from the external entities are translated by this layer to create the objects in the Core layer. Three kinds of inputs are provided to the broker: a description of the application requirements, a set of services that can be utilised for executing the application, and the set of credentials for accessing the services.

The application description provides details of the execution such as what is to be executed, the description of task inputs including remote data files (if required), the information about task outputs (if present) and the desired quality of service. This description can be provided in one of the XML-based languages supported by the broker, or via the broker's APIs. Similarly, the set of services required for the user objectives can be provided through the APIs or as an XML-based service description file containing information such as service location, service type and specific details such as remote batch job submission systems for computational services. The services can also be discovered by the broker at runtime from remote information services such as the Grid Market Directory (GMD) [17] or Grid Index Information Service (GIIS)[18] among others. The list of credentials is provided in another file. File-based inputs are handled by the respective interpreters which convert the descriptions to entities within the broker. The Application Interpreter converts the application description file to Task objects while the Service Interpreter converts the service description to Service objects. The Credential Interpreter similarly creates Credential objects that are associated with a user and used by the broker to authenticate while invoking services (e.g. dispatching jobs on computational resources). These objects are described as a part of the Core layer in the following section.

## 2.2 Core Layer

This layer contains entities that represent the properties of the Grid infrastructure independent of the middleware and the functionality of the broker itself. Therefore, it abstracts the details of the actual interaction with the Grid resources performed by the Execution layer. This interaction is driven by the decisions made by the functional components of the broker present in the Core layer. These components can be broadly classified into two categories - *entities* and *workers*. This terminology is derived from standard business modelling processes [19].

Entities exist as information containers representing the properties, functions and instantaneous states of the various architectural elements that are proxies for the actual Grid entities and constructs involved in the execution. Therefore, entities are stored in the persistence database and are updated periodically. Workers represent the functionality of the broker, that is, they implement the actual logic and manipulate the entities in order to achieve the application objectives. Therefore, workers can be considered as active objects and the entities as passive objects. The next section (Section 3) takes a closer look at the entities and workers within the broker accompanied by UML 2.0 [20] diagrams that illustrate the relationships between the components.

## 2.3 Execution Layer

The actual task of dispatching the jobs is taken care of by the Execution layer which provides Dispatchers for various middleware. These dispatchers create middleware-specific Agents from the jobs and are executed on the remote resources. If there are any data files associated with the job, then the agents request them from the data repositories that have been selected to access those files. During execution, the Job Monitor keeps track of the job status - whether a job is queued, executing, has finished successfully or has failed on the remote resource. On completion of job execution, the associated agent returns any results to the broker and provides debugging information.

## 2.4 Persistence Sub-system

The persistence subsystem extends across the three layers described previously and maintains the state of the various entities within the broker. It is primarily used to interface with the database into which the state is stored at regular intervals. The persistence sub-system satisfies two purposes: it allows for recovery in case of unexpected failure of the broker and is also used as a medium of synchronisation among the components in the broker.

# 3 Design of the Gridbus Broker

This section presents the design of the Gridbus broker in detail. As mentioned before, the components in the broker are termed as entities or workers depending on whether they are passive or active objects. We will take a closer look at each of these, and later will explain the how the design was driven by the requirements mentioned in Section 1.1.

## 3.1 Entities

*Application Context and Tasks*

An `ApplicationContext` represents the attributes of a user's application specification such as the description of the application to be executed, one or more credentials for accessing the services for performing the execution and QoS requirements codifying user's expectations of the execution such as a deadline by which the execution must be completed. Figure 3 shows the placement of this object in relation to tasks and jobs. The `ApplicationContext` provides a single point for accessing a user's requirements.

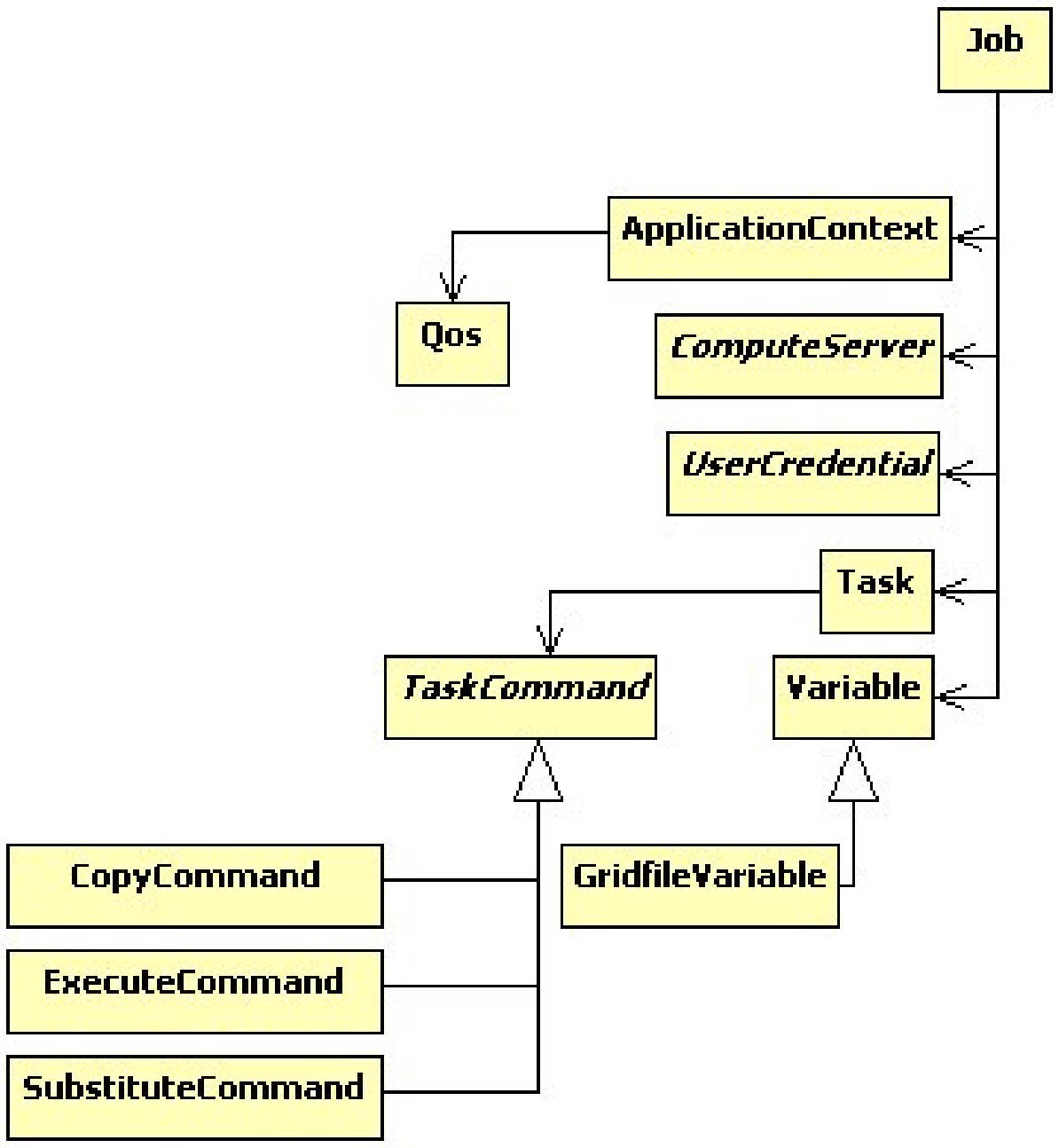

*Figure 3: ApplicationContext, Tasks and Jobs.*

A `Task` object represents the activities to be carried out in order to satisfy the user's requirements. Examples of activities include copying a file to the remote node, running the executable and storing the output in a remote repository. A `Task` consists of multiple `TaskCommands`, each of which encodes an activity, and which are carried out in sequence. Figure 3 illustrates three activities which inherit from the task command – the `CopyCommand` object which represents a file transfer between two machines, the `ExecuteCommand` object which specifies the executable and its arguments and the `SubstituteCommand` that specifies files within which variable values have to be substituted at runtime. The `TaskCommand` structure makes it possible to add more types of activities as and when there are requirements from applications.

Tasks are associated with input parameters encoded in `Variable` objects described later. The task specification provides the template for creating jobs which are the actual units of work sent to the remote Grid resources. That is, a task is an abstraction of the work performed for executing the application. The conversion of task to jobs is scenario-dependent; for example, a parameter sweep task is converted into a set of jobs, while a single, independent task representing a simple application with a single function is converted into one job. The task structure within the broker is based on and extended from the task specification followed by Nimrod[21] for parameter sweep applications. This task structure was chosen as it was generic, simple and could be mapped to any task description language. For example, the broker supports two different task specification languages: XPML (eXtended Parametric Modelling Language), an XML-based language created by us based on Nimrod's "plan file" format for describing parameter sweep applications; and JSDL (Job Submission Description Language)[22], an Open Grid Forum (OGF) standard for describing independent batch job submissions.

### *Job*

A `Job` represents an instantiation of the task at the remote node and is therefore associated with a single Task object that describes its function. The structure of the `Job` object is shown in Figure 3. A Job may be associated with one or more `Variable` objects which describe its input data. A `Variable` can be one of various types including integer, string and float and is associated with a single value derived from its domain. New types of variables can be introduced into the broker by extending the `Variable` class. For example, it was extended to create a `GridfileVariable` which describes a file or a dataset that is stored on a repository that is accessible through any of the supported file transfer protocols.

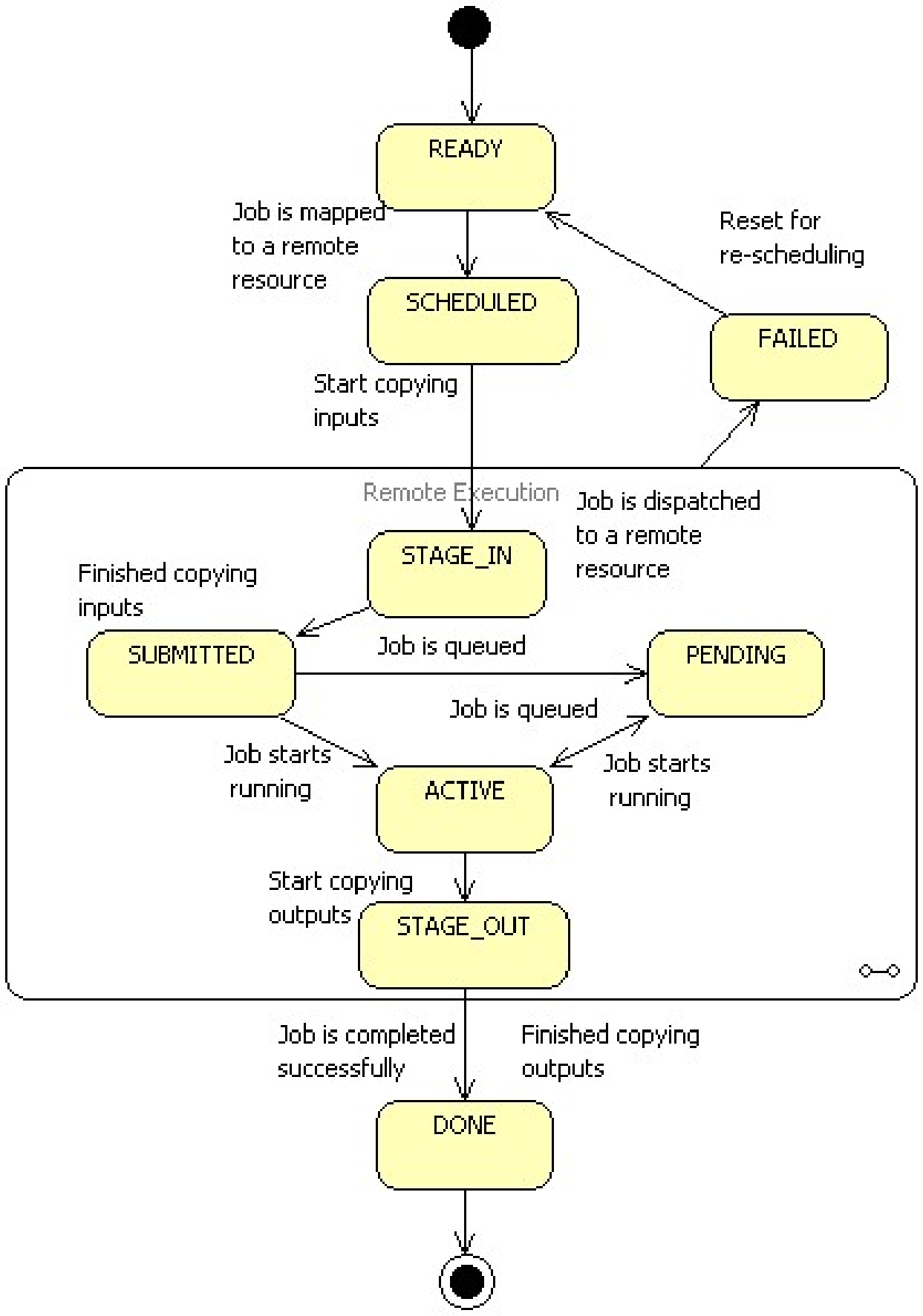

*Figure 4: State transition diagram for a job.*

A Job is also associated with a JobWrapper that represents the interface for translating the user task specification to create an Agent that can be executed by the middleware running on the designated compute resource. Therefore, the JobWrapper is necessarily middleware-specific and is associated with the Execution Layer. We will discuss more about the JobWrapper in relation to scheduling and job dispatching. Other than these, the Job is associated with the ApplicationContext of its Task specification, a set of services to which it is allocated and a UserCredential.

In the course of its lifetime, a job passes through many states as is outlined in Figure 4 . A job is an input to the scheduler which allocates it to a set of resources based on its requirements. The job's status is then changed to SCHEDULED. During the STAGE_IN state, input files and executables required for the job are staged to the remote resource. When this process is completed successfully and a handle is obtained, then a job is considered to be SUBMITTED. The job may be queued while waiting for an available processor and its state changes to PENDING. When the job starts its execution, it is considered ACTIVE. After the job has finished executing, it enters the STAGE_OUT stage where its output files are transferred back to the broker. If all its outputs are received and are as expected by the task requirements, then the job is considered as "DONE". If one of state transitions fails on the remote side or the job has completed on the remote side but has not produced the expected result files, then it is considered FAILED and is reset and marked for re-scheduling.

### *Services*

It is possible to represent different types of services within the broker mirroring the variety of services that are available in Grids. These are represented in a hierarchy, shown in Figure 5 in which the parent class is an abstract Service whose attributes are the Service ID and the location of the service (hostname or Uniform Resource Indicator (URI)). The next level groups the services into categories depending on the type of the service. The ComputeServer object represents a computational resource with properties such as architecture, operating system and available job submission systems. DataHost objects describe storage repositories and the details of the data files stored within these. These details include attributes such as the location, size and the protocol used to access the files in the repository. Services that provide meta-information such as resource information services, data catalogues and market directories are represented by InformationService object. InformationServices are categorised depending on the type of information they serve. For example, ReplicaCatalog services such as Globus Replica Catalogue and SRB MCAT provide information on different copies of required datasets that are stored on distributed repositories. Information about network properties is gathered from the NetworkInformationService and is stored in data structures called NetworkLinks that keep track of the changing network conditions between various resources.

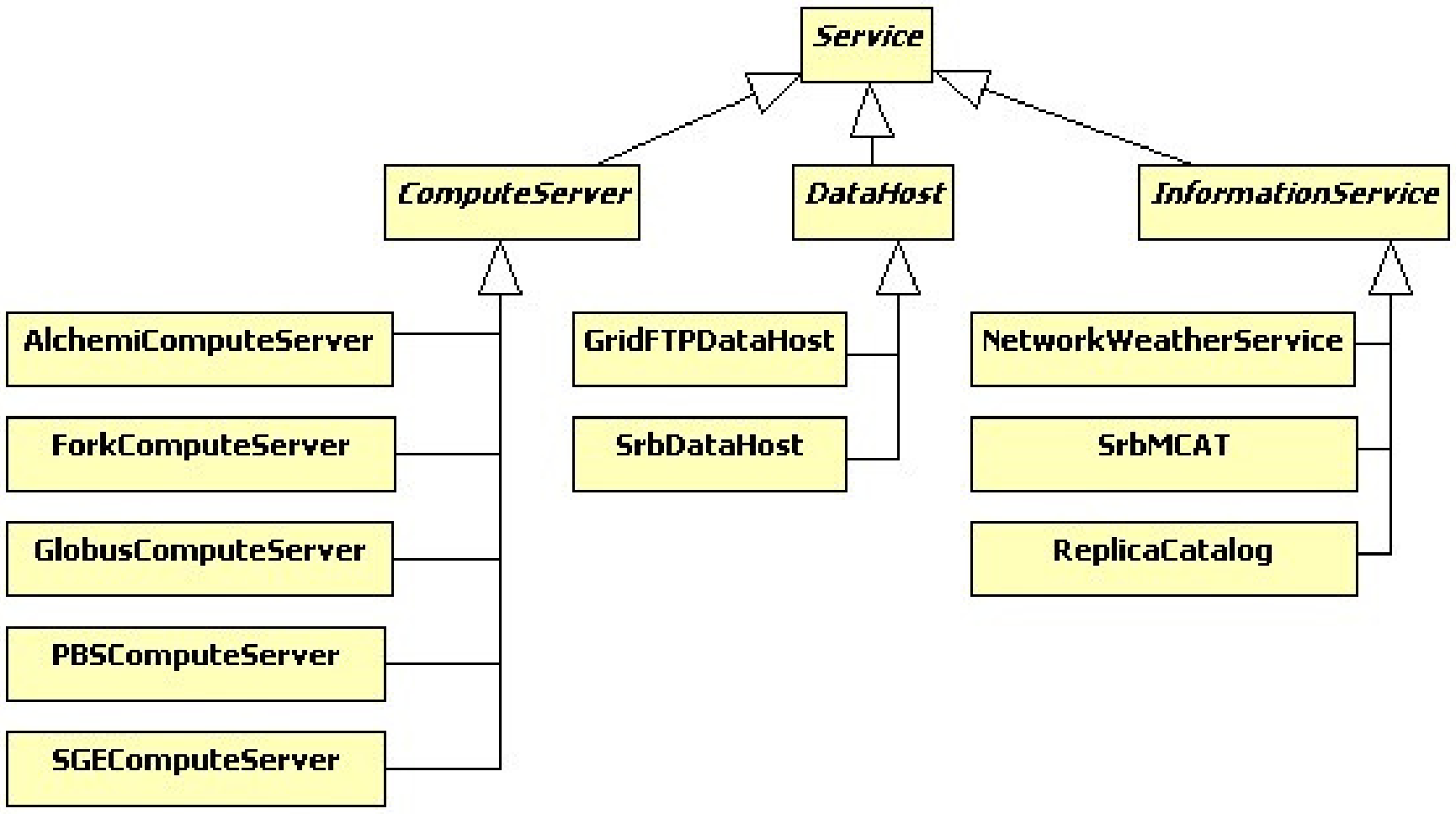

*Figure 5: Service object hierarchy.*

The lowest child classes in the hierarchy represent specific implementations of these services that provide the functions of associated middleware. For example, interaction with a computational resource running Globus middleware is implemented in the `GlobusComputeServer` class that extends the abstract `ComputeServer`. This interaction, which utilises the Java Commodity Grids (Java CoG) Kit in the Execution Layer, is kept isolated from the rest of the broker to which the `ComputeServer` class is presented as a single compute resource. In production Grids, it is more likely that the remote resource may be a cluster running some sort of a batch job management system such as Portable Batch Scheduler [26] or Sun Grid Engine (SGE) [27] . These job queueing systems may be accessible either through Globus job managers or directly using command line utilities. Many of these systems are set up to have queues with different priorities, number of available slots and limits on the duration of jobs. The broker supports such systems by providing facilities for scheduling jobs to queues based either on the user's specification or on the length of the job and the availability of slots. Similar to Globus, the abstract `ComputeServer` has also been implemented for different middleware and job managers such as Alchemi[24], Unicore[25], PBS, SGE, Condor[28] and XGrid[29]. In the case of the last four, the job management systems are invoked through an SSH (Secure Shell) connection.

```
package org.gridbus.broker.services.data;

public class SrbDataHost extends DataHost {

        public String getMdasDomainName();
        public void setMdasDomainName(String domain);
        public int getPort();
        public void setPort(int srbPort);
        public String getUsername();
        public void setUsername(String username);
        public void setPassword(String pwd);
        public String getPassword();
        public String getAuthScheme();
        public void setAuthScheme(String auth_scheme);
        public String getMdasResourceName();
        public void setMdasResourceName(String mdasResourceName);
        public String getZone();
        public void setZone(String zone);
        public String getServerDN();
        public void setServerDN(String serverDN);
}
```

*Figure 6: List of methods in SrbDataHost.*

Interaction with remote storage resources is implemented through the `DataHost` abstraction. Specific implementations of this abstraction provide specific functions of storage middleware. As an example, Figure 6 shows the list of methods available in the `SrbDataHost` class that implement functionality specific to Storage Resource Broker (SRB) such as Zones and SRB authentication. Similarly, the `DataHost` has been implemented for providing access to data stores Grid-enabled by GridFTP.

### *Credentials*

The Gridbus broker defines the `UserCredential` object that represents an authentication token to access remote services. The base `UserCredential` object has been extended to realise the different types of credentials that are accepted by different middleware. For example, the `LocalProxyCredential` represents the Globus GSI (Grid Security Infrastructure) X.509 proxy object created on the client side. On the other hand, a `SimpleCredential` is used for accessing resources that require a username and a password for authentication such as those enabled through SSH (Secure Shell).

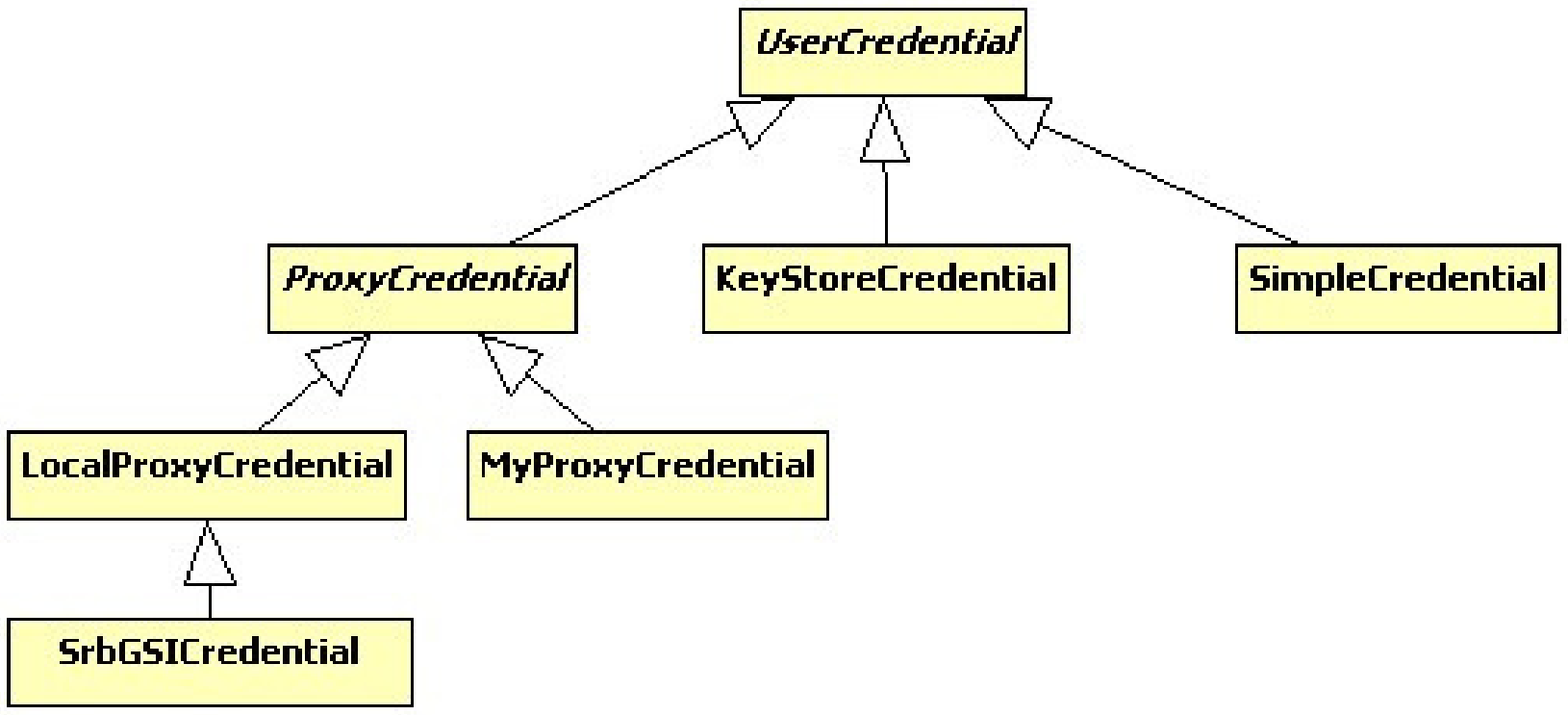

*Figure 7: Credentials supported by the broker.*

The set of credentials available to the user is associated with his/her `ApplicationContext`. Even though the credentials are passive objects, to ensure the security of users' credentials, these are transient and are not saved in the persistence storage. This also means that the user has to provide the broker with a fresh set of credentials when recovering from a previously paused/failed run of a Grid application. Figure 7 shows the various types of credentials supported by the broker.

## 3.2 Workers

*Broker*

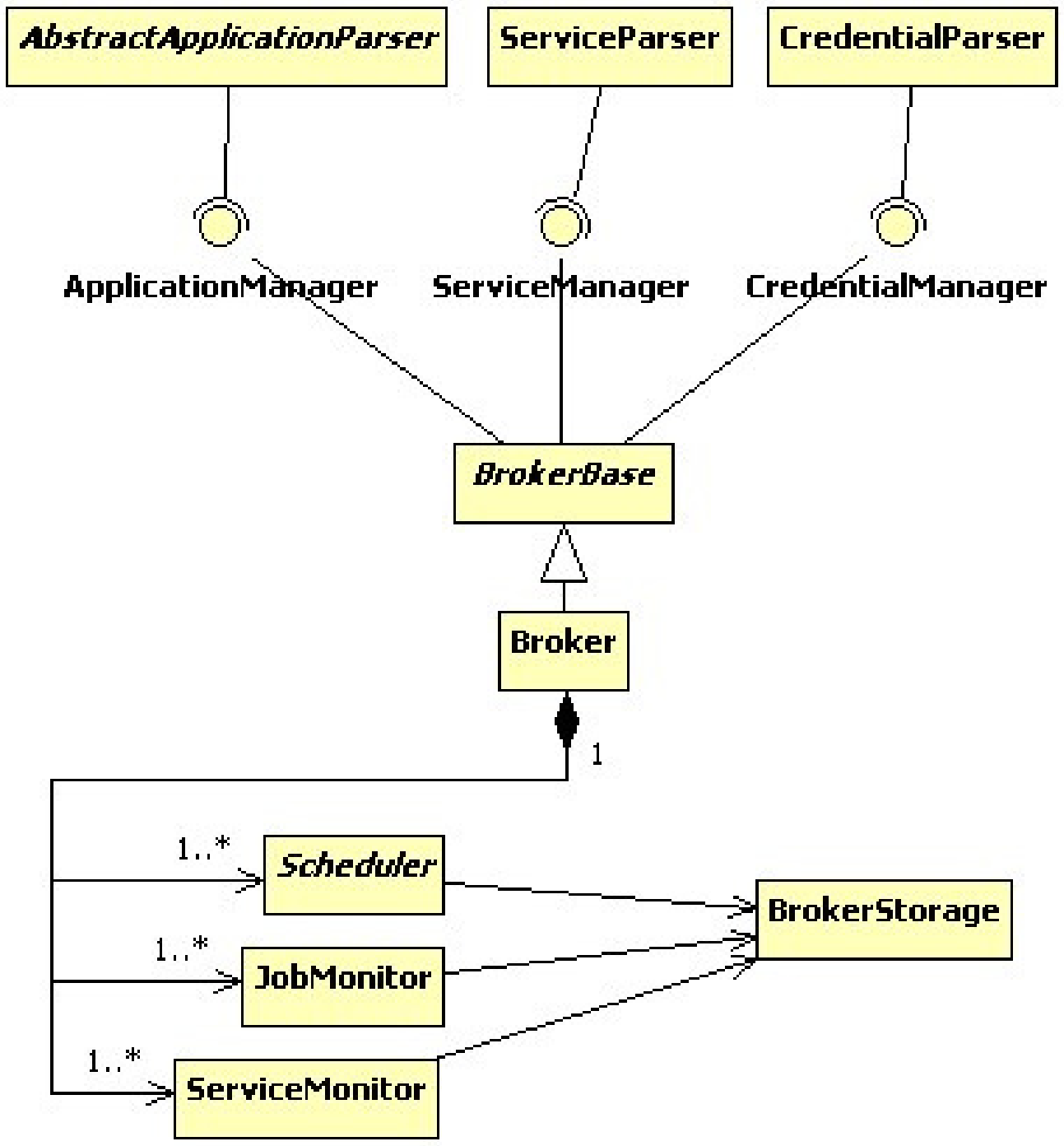

*Figure 8: The Broker and its associated entities.*

The `Broker` object is the first component to be initialised and is a container for the other objects in the broker. It is responsible for managing the lifecycle of the other active objects – `Scheduler`, `JobMonitor` and `ServiceMonitor` – from start-up to shutdown. It is also a front end to the persistent storage and therefore, maintains the overall state of the broker by saving the state of various passive entities. The Broker and its associations with other entities within the broker are shown in Figure 8. The BrokerStorage class provides the methods for the other active objects to interact with the persistent database. The Parser objects convert the various input specifications to the entities within the broker and store them in the database.

The broker can be started as a single-user system on the command-line, a remotely hosted web service compliant with the WSRF (Web Service Resource Framework)[30] standard or embedded within another Java application. The WSRF interface allows the isolation of the Broker functionality from the hosting application. A common usage scenario for this mode of operation is a web portal that offers a front-end to e-Science applications. The WSRF interface for the broker is designed as an adaptor around the Broker component. The WSRFBroker is a wrapper for the broker class which parses the SOAP messages and delegates all the invocation to the broker internally.

### *ServiceMonitor*

The ServiceMonitor component is responsible for keeping the broker's view of the grid up-to-date. It periodically checks the availability of the specified remote services and discovers new services which may become available. It is able to retrieve information from various Service objects about the grid environment. The information retrieved is stored into the broker's database from where it is retrieved by the scheduler for making decisions.

```
/**
* Checks if the compute server is up, and sets all its attributes
* @return true if the properties have been discovered
*/
protected boolean discoverProperties(UserCredential uc) {
   try{
            ProxyCredential pc = (ProxyCredential)uc;
            . . . . . .
            // Building the query string
            String filter = "(&(objectclass=MdsHost)(Mds-Host-hn="+this.getHostname()+"))";
            // Querying the remote Globus Resource Information service
            NamingEnumeration results=MDSUtil.search("ldap://"+this.getHostname()+":2135",
               filter, HOST ATTRIBUTES);
            if(results==null) {
                logger.error("setValues() - Error in accessing MDS!!" ,null);
            } else {
                while(results.hasMore()) {
                  // Set the attributes in GlobusComputeServer
                  . . . . . .
            }
  . . . . . .
    return true;
}
```

*Figure 9: The implementation of Service.discoverProperties() in GlobusComputeServer.*

Initially, it polls all available services by invoking the discoverProperties operation that is provided within each Service entity. This operation does a search for service attributes that are specific to the service and middleware type. If the operation is successful, the values that are so retrieved are set within the Service object and the Monitor is notified that the service is available. The aim of this is to allow each Service to define its own method of obtaining attributes. An example of the implementation of discoverProperties for GlobusComputeServer is shown in Figure 9. This implementation discovers attributes of resources running Globus 2.x middleware. It first checks if the remote resource is alive and if so sets up a filter with the required attributes and performs a query. The results of the query are then stored in the respective fields within the object. If the query fails, then the attributes will have null values. However, query failure is not escalated to the rest of the broker.

### *Scheduler*

The scheduling component is designed as two separate components: the Scheduler and the Dispatcher. The Scheduler matches the jobs individually to the services and also, decides the order of execution of the jobs on the resources. Figure 10 shows the basic sequence of operations that is performed by the `Scheduler`. The `Scheduler` gets the list of ready jobs from the persistent storage and a list of services, depending on the strategy, that it is interested in. The mapping is an assignment of a job to appropriate computational and data services. The job can specify minimum requirements for a qualifying resource. The `Scheduler` takes this into account and uses appropriate heuristics (based on user selection) to generate the mapping. At the very least, the job has to be mapped to a compute resource (or a `ComputeServer` object) where it is to be executed. If the application specifies data files to be processed, then the mapping also includes the assignment of `DataHost` objects from which each of the files should be accessed. At the same time, the state of the Job is changed from READY to SCHEDULED.

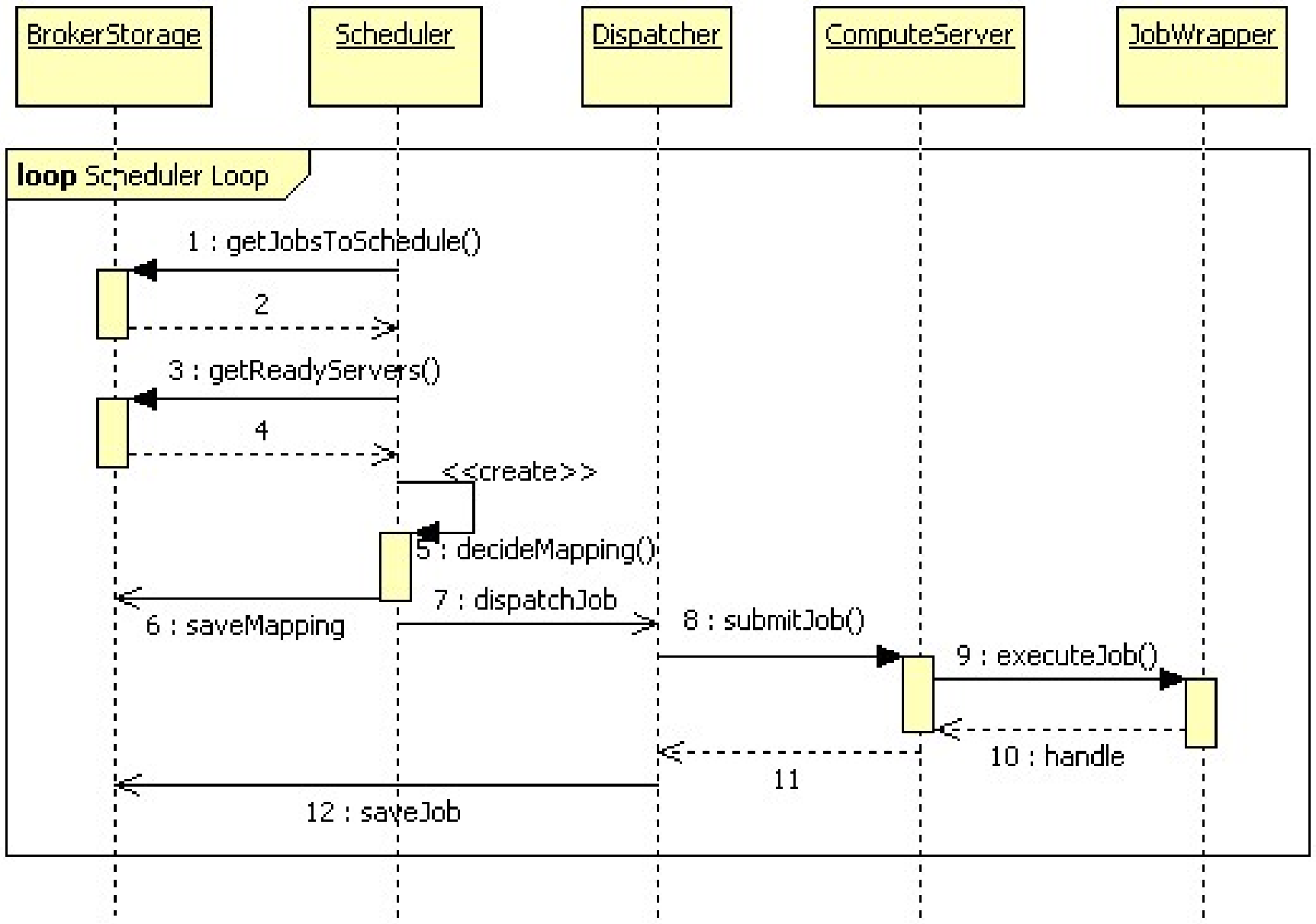

*Figure 10: Schedule sequence diagram.*

The `Dispatcher` created by the `Scheduler`, to dispatch a job creates a `JobWrapper` object depending on the type of `ComputeServer` selected for the job. The `JobWrapper` is a short-lived object which translates the abstract Job to execute on the target middleware and architecture. For example, the `Dispatcher` would instantiate a `GlobusJobWrapper` object for a `GlobusComputeServer` which translates the `TaskCommands` to invocations specific to the target architecture and Globus 2.x. It then performs job submission in accordance with the protocols followed by Globus.

Prior to the actual submission of the job, the files required for the job such as input files and executables (if any) are copied on to the remote resource. The files can be transferred on to the resource by the broker (push model) or they can be requested and copied by the remote resource (pull model). This allows a great deal of flexibility in implementing transfer modes, three of which are illustrated in Figure 11. For example, many Grid resources are behind firewalls that prohibit any connections to any port on the resource. In this case, using the pull model has the advantage that a transfer program on the resource can make an outbound connection through the firewall (Figure 11 (b)). Another advantage of the pull model is that the source of the files can be a file server that is separate from the resource on which the broker is running thereby supporting a scenario in which the broker is behind a firewall as well (Figure 11 (c)). The resource manager is a middleware component that is able to send and receive messages from the outside world through the firewall. During this process, the Job state is changed to STAGE_IN.

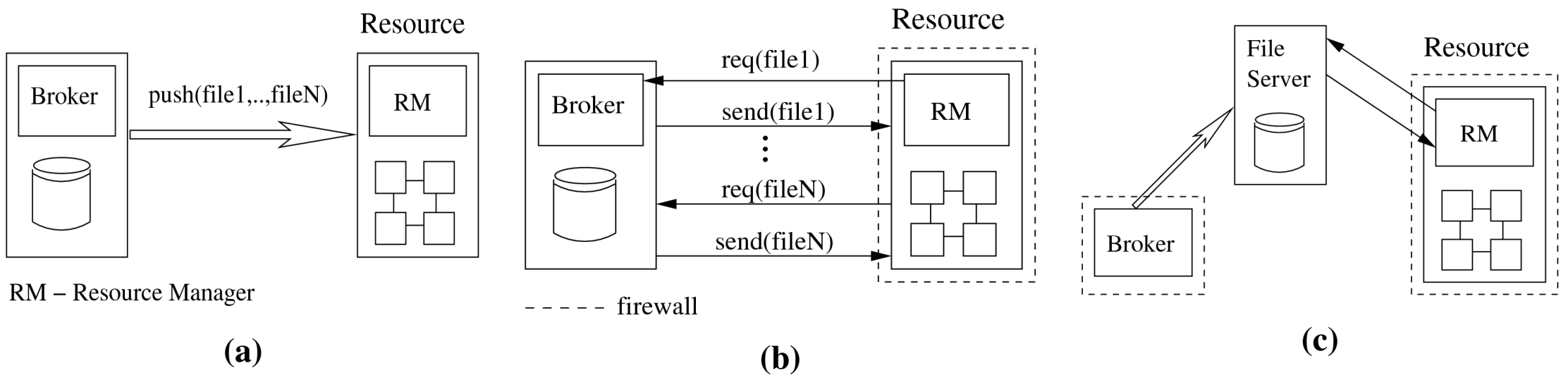

*Figure 11: File transfer modes supported by the broker. (a) Push model of file transfer. (b) Pull model of file transfer. (c) A file transfer using an intermediate server.*

The `JobWrapper` then submits the prepared job to the remote resource manager and waits for confirmation of acceptance. This is obtained as a remote handle to the job that uniquely identifies the job at the resource. The job state is changed to SUBMITTED and the job along with the handle is saved back to the persistent storage. The number of available slots within the `ComputeServer` is decremented to reflect the submission and the server also saved to the persistent storage. If the handle is not received, then the job submission is considered as FAILED and the job is marked for re-scheduling.

The `Dispatcher` is also able to follow two-phase commit protocol for job submission as described by Czajkowski et al. [31]. In the first phase, the stage-in of the files and the job submission are performed and the dispatcher waits for an “agreement" message from the remote resource in the form of the remote handle. After the remote handle is received, the dispatcher sends a “commit" message to the remote resource which proceeds with the job submission and sends an acknowledgement back. The job state is changed to SUBMITTED only after the receipt of the acknowledgement. For resources running middleware that do not support the two-phase protocol, the receipt of remote handle is considered as acknowledgement of submission. In this case, if there is a network failure before the handle is received at the broker, the job will be marked as failed though it may have started execution at the remote node. In two phase protocol, the job is not processed by the Grid resource until the broker sends a commit message. Thus, two-phase commits ensure that job submission is carried out only once even in the face of network problems.

Prior to passing control to the `JobWrapper`, the `Dispatcher` selects a credential from the set of `UserCredentials` and binds it to the job. This decision is based on the type of the middleware and the type of credential. Alternatively, a user may explicitly specify the mapping of credentials to resources. This feature aids a user to seamlessly run jobs on different types of middleware, or even to use different credentials for the same type of middleware, at the same time.

### *JobMonitor*

The `JobMonitor` object is associated with the `ComputeServer` keeps track of a job's progress after submission to that remote resource. As shown in Figure 12 , the `JobMonitor` periodically (the default is 30 seconds) requests the list of jobs that are in the SUBMITTED state on a particular resource, from the persistent storage. It uses the remote handle to query the status of the job using middleware-specific functionality. The query is a blocking call, and is therefore provided with a timeout period after which the `JobMonitor` cancels the query to proceed to the next job. Failure to contact the job on the remote resource is not considered as a failure of the job immediately. The `JobMonitor` tries to contact the job again for a set number of times before giving up and marking the job as failed.

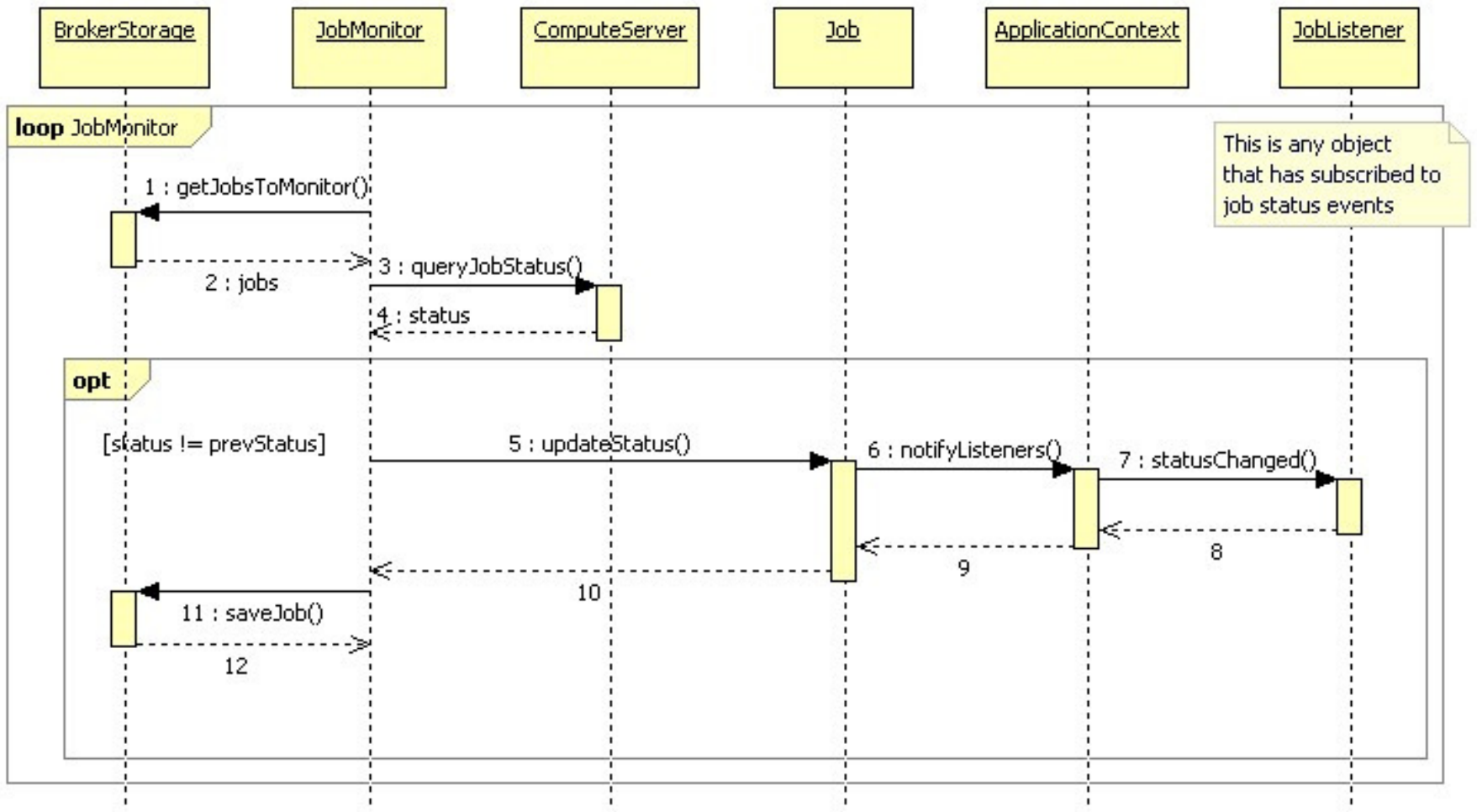

*Figure 12: Job monitoring sequence diagram.*

It is possible to implement `JobListeners` interfaces that will receive events from the `JobMonitor` when a job status is changed. The listeners can be entities within the broker such as an event-driven scheduler or outside the broker such as an applet within an application Web portal that has been integrated within the Gridbus Broker [32].

## 3.3 Meeting Challenges in the Design

The primary aim of the Gridbus broker is to provide a generic resource discovery and scheduling framework that abstracts the heterogeneous and dynamic nature of the Grid infrastructure and allows users to achieve different objectives. In Section 1.1, some of the challenges in achieving this aim are outlined. These challenges provide the requirements against which the broker is designed. The broker meets these requirements in the manner outlined as follows.

### 3.3.1 Service Heterogeneity

The Gridbus broker tackles the problem of heterogeneous Grid resources and service interfaces by adopting the principle of *minimal assumptions*. This means that throughout the broker, there are as few assumptions as possible about the nature of the environment in which it operates. The relationship between the objects in the broker is generic and independent of interaction models followed by any middleware. The broker, therefore, does not impose a particular configuration requirement on Grid resources and is thus able to use as many resources as possible. For example, a resource running any Unix-based operating system and with one of the supported middleware operational would be immediately useable by the broker as the latter requires only a POSIX-compliant shell environment that is standard on such machines.

The three-layer architecture of the broker also helps in maintaining this independence. For example, the assignment of jobs to resources is performed by the Scheduler which, as a component of the Core layer, has a middleware-independent view of the resources, while the Dispatcher dispatches the jobs to the resources. However, the actual interface with the remote resource happens through the middleware-specific JobWrapper which is a part of the Execution layer.

The broker can also be made to interface to any new service or middleware by extending the appropriate classes. For example, support for a new middleware can be added by extending the abstract ComputeServer and JobWrapper classes. The Scheduler will then be able to immediately utilise any resource running that middleware. The interaction with the remote middleware is also independent of the rest of broker. Using this method, the broker has been extended to support a wide variety of both computational and data Grid middleware such as Globus, PBS and SRB among others [33][34]. With the support for Alchemi, which is a .NET-based desktop Grid computing framework for Windows platforms, and XGrid, a similar framework for Mac OS X, the Gridbus broker can already schedule jobs across almost all of the platforms in use today. Similarly, new information sources and data repositories can be supported by extending the InformationService and the DataHost classes respectively.

### 3.3.2 Support for Different Application Models and User Interfaces

The components within the broker are designed to be modular and there is a clean separation on the basis of functionality. Particularly, the division between workers and entities clearly delineates responsibilities among the components of the broker. Since the passive entities are just holders of information, the logic within the workers can be changed without affecting the former. It is also possible to introduce new workers that either use or extend the existing entities in different ways or introduce new entities of their own *without structural or design changes to the rest of the broker*. This loose coupling between components provides a lot of flexibility and enables the realisation of different application and system models [35].

The components within the Interface Layer convert the application, service and credential descriptions to broker entities and store them into the persistence database. Thus, it is possible to support any form of description by mapping it to the entities within the broker. At present, the broker supports XML-based description of parameter sweep and bag-of-task applications and has a set of XML-based files for describing services and credentials. These inputs can also be provided directly to the broker through its APIs. The APIs allow direct manipulation of the entities in the broker and thus, can be used to realise different application models.

The same mechanisms that allow the creation of application models also enable the creation of different user interfaces for the broker. The broker can be interfaced through command line, desktop clients or web portals. It is also possible to talk to the broker's persistence database directly through its APIs. This enables any component to retrieve information about the broker entities through normal SQL (Structured Query Language) queries. This ability is useful in scenarios such as an application Web portal requiring information about an ongoing execution through a broker installed on a machine different from the portal server.

The WSRF web service interface enables multi-user multi-application scenarios, where the broker is hosted as an isolated service that is accessible by web portals using the WSRF Broker Client APIs. The separation between web application and broker service reduces the memory and CPU usage in the web server hosting the application, and increases performance and scalability. Moreover, the reliability of the application is also enhanced as failures in the broker will not affect the application server, and end users could still have the application web portal working while the broker service recovers and resume execution of the grid application.

### 3.3.3 Realisation of Different User Objectives

The broker allows for schedulers to be plugged-in, and hence is able to support new scheduling algorithms and policies. This enables the broker to adapt to new user requirements and objectives. The separation of the dispatch component from the scheduling provides a lot of flexibility for implementing the scheduling logic. These two components are also designed to be independent of the resource and middleware details. However, a developer can still choose to create schedulers that may require certain middleware-dependent services such as resource information services.

One of the main requirements of the design was the ability to execute generic data-oriented applications that may require access to one or more distributed datasets at the same time. Data services, represented by `DataHosts`, have the same level of importance as `ComputeServers`. Location of available replicas of a dataset can be gathered by querying the appropriate replica catalogs that are represented as `InformationServices`. Information about current network conditions such as available bandwidth, are important when large datasets are to be transferred. This is available through the `NetworkLinks` data structure as are properties such as pricing, classes of service and availability. `CopyCommands` in the Task allow the application to perform third-party (not involving the broker machine) point-to-point data transfers on the Grid. The application interpreter allows users to specify datasets as input parameters to their applications. In this case, the `ServiceMonitor` will discover all the data repositories that host these datasets (through implementations of data catalog information services) and the `Scheduler` will select one of the repositories for accessing the datasets.

The Gridbus broker has been designed from the ground up to support the computational economy paradigm [36], and assigns costs to various services including computation, data storage, and information services. The default scheduling policy uses these parameters to decide on an optimal mapping strategy to schedule jobs on resources. The design also allows the ability to plug-in a market directory which offers information about various priced services, and a Grid bank that manages users' credit in a Grid market. The

`ServiceMonitor` is able to periodically refresh pricing information from market information services and therefore, provides the ability to keep up with dynamic pricing and resource conditions in a volatile Grid economy. More importantly, the combination of all these features enables the broker to satisfy different requirements such as selecting data repositories for accessing datasets on the basis of price and/or performance and selecting network links offering a particular class of service[37].

The Gridbus broker is also able to support users who would like to use resources spread across multiple VOs by natively managing multiple credentials for multiple resources. This ability also allows users to access legacy data repositories that are not “Grid-enabled" and follow their own authentication mechanism, rather than Grid proxies.

### 3.3.4 Infrastructure

Grid environments are dynamic in nature. As a result, transient behaviour is not only a feature of the resources but also of the middleware itself. The broker design considers various types of failures that may occur either on a remote resource to which a job is submitted or on the broker machine itself during various stages of its operation. Remote failures include failure during job submission, execution and monitoring, or retrieving the outputs. These could be due to various reasons, such as incorrect configuration of the remote resource, incorrect job descriptions, network problems, unavailable data files, or usage policy restrictions on the resource. Local failures include unexpected system crashes which lead to abrupt termination of the broker, unavailability of local input files and invalid input parameters.

The broker applies different fault recovery methods in different cases of remote failure. A job is not considered completed until it is determined that each of its constituent task activities has successfully exited. Thus, even if the middleware on the remote resource has signalled a successful completion, the JobMonitor checks to see if the job has generated required results and only then it is deemed successful. The JobMonitor may not be able to contact an executing job in case of transient network conditions. In such cases, the JobMonitor polls the job a set number of times, and if it is able to re-establish contact, then check the job’s current status. In all other cases of job failure, the job is rescheduled on another available resource.

The persistence system provides insurance against failure of the broker itself. At any point in the execution, the state of the broker is completely described by the contents of the entities and therefore, only these need to be stored within the persistence database. Any change in the status of a Job or a Service, as discovered by the respective Monitors, is immediately written back to the database so that the latter reflects the true state of the broker. Workers have no state of their own and hence, can be resumed from the point at which they failed by reading the entities from the persistent storage.

The broker tracks variations in resource availability by monitoring resource performance locally. Nothing is required to be installed at the remote resources and dependencies on metrics provided by the middleware are avoided by default. Thus, the broker is able to compare resources in a heterogeneous environment based on metrics that are independent of middleware and relative to the requirements of the current execution. However, it is to be noted there is no mechanism in the broker to inhibit usage of external performance monitors and other services if required.

## 4 Related Work

The challenges presented in Section 1.1 have motivated the development of a large number of Grid resource brokering and application deployment systems. Examples of such systems are Nimrod/G, Condor-G, APST, EU-DataGrid Workload Management System (later succeeded by the gLite[39]), GridWay[40], DI-GRUBER[41], SPHINX[42], and GridLab Grid Resource Management System[43] among others. In this section, however, the first four are chosen for detailed comparison against the Gridbus broker as their objectives and approach are similar to that of the broker. These are compared to the broker against the manner in which they handle the challenges outlined previously.

### 4.1 Nimrod/G

Nimrod/G[9][10] is a tool for automated scheduling and execution of parameter sweep applications on Grids. It provides a declarative parametric modelling language through which the task specifications can be provided for an “experiment" or execution of an application. Scheduling within Nimrod/G follows an economic model in which the resources have costs associated with them and the users have to expend their budgets in order to execute their jobs on the resources. The user can also specify Quality of Service (QoS) requirements such as a deadline for finishing the experiment and an option for choosing between a faster yet more expensive execution vis-a-vis a slower but cheaper process. Architecture-wise, Nimrod/G consists of a Task Farming Engine (TFE) for managing an execution, a Scheduler that talks to various information services and decides on resource allocations, and a Dispatcher that creates Agents and sends them to remote nodes for execution. An Agent can manage more than one job at a remote site.

Nimrod/G works with UNIX-based resources enabled through Globus, Legion, PBS, Condor, SGE and others. It has the ability to specify data transfers between any two machines that may be running different Grid middleware. It takes into account different patterns of data usage such as pipelined data between sequential processes to schedule jobs and to make better use of storage on resources [56]. Nimrod/G proposed the Grid economy paradigm and has implementations of four algorithms [52] - *time optimisation*, *cost optimisation*, *cost-time optimisation* and *conservative time optimisation* - for scheduling parameter sweep computationally-intensive applications. It also implements a number of other scheduling algorithms from APST, leveraging services such as the Network Weather Service. Nimrod/G also provides additional interfaces that expose the functionality via Web Services and Web portals, Moreover, tools like Nimrod/O[57] allows users to specify complex search and optimization algorithms.

## 4.2 AppLeS Parameter Sweep Template (APST)

APST[11] is an environment for scheduling and deploying large-scale parameter sweep applications (PSAs) on Grid platforms. APST provides mechanisms for deploying applications on different Grid middleware and schedulers that take into account PSAs with data requirements. APST consists of two processes: the daemon, which deploys and manages applications and the client, which is a console for the users to enter their input. The input is XML-based and no modification of the application is required for it to be deployed on Grid resources. The APST Scheduler allocates resources based on several parameters including predictions of resource performance, expected network bandwidths and historical data. Examples of scheduling strategies include algorithms that take into account PSAs with shared input files [48] and Divisible Load Scheduling-based algorithms [49]. The scheduler uses a Data Manager and a Compute Manager to deploy and monitor data transfers and computations respectively. These in turn use Actuators to talk to the various Grid middleware. A Metadata Manager talks to different insformation sources such as Network Weather Service (NWS) [51] and the Globus Monitoring and Discovery Service (MDS) and supplies the gathered data to the scheduler.

APST supports different low-level Grid middleware through the use of Actuators and also allows for different scheduling algorithms to be implemented. However, it is focused towards parameter sweep applications. APST provides the ability to specify data repositories of different types in the input file and has a separate data manager to manage data transfers.

## 4.3 Condor-G

Condor-G[12] is a computational management system that allows users to manage multi-domain, heterogeneous resources running Globus[42] and Condor[45] middleware, as if they belong to a single domain. It combines the harnessing of resources in a single administrative domain provided by Condor with the resource discovery, resource access and security protocols provided by the Globus Toolkit. At the user side, Condor-G provides API and command line tools to submit jobs, cancel them, query their status, and to access log files. A new Grid Manager daemon is created for each job request which then submits the job to the remote Globus gatekeeper that starts a new JobManager process. Condor-G provides “execute once" semantics by using a two phase commit protocol for job submission and completion. Fault tolerance is provided on the submission side by a persistent job queue and on the remote side by keeping persistent state of the active job within the JobManager. Jobs are executed on the remote resource within a *mobile sandbox* that traps system calls issued by the task back to the originating system and are check-pointed periodically using Condor mechanisms. This technology called Condor GlideIn effectively treats a collection of Grid resources as a Condor pool. Resource brokering is provided by matching user requirements with information available from services such as GRIS and GIIS through the ClassAds[46] mechanism.

Condor-G operates in a Globus and Condor-only environment and installs a virtualization layer at each node at runtime that traps system calls and provides check-pointing facilities. Condor can utilise batch queuing systems such as LSF, PBS and NQE but only through Globus GRAM protocols. Condor-G provides strong fault tolerance mechanisms as a result of its close integration with the low-level Grid middleware. It implements the two-phase commit protocol for Globus job submission for ensuring that the job is executed only once. Through the GlideIn mechanism, it is able to provide libraries that perform check-pointing and job migration and maintains a persistent queue to guard against local failures.

Though Condor-G by itself does not provide any data access functions, it can interface to services such as Kangaroo[47] and Stork[48] that enable it to mediate access to remote files and manage data transfers. Condor-G allows for creation of applications belonging to different models such as workflows and supports different scheduling strategies. However, it does not natively support resource costs and has no functions for optimisations based on pricing.

## 4.4 gLite Workload Management System

gLite[39] is an integrated middleware package for the EGEE project that consists of modules for security, information and management, data and job management services. Here the focus is on the gLite's WMS (Workload Management System) package that provides access to resources running various middleware such as Globus, Condor and Storage Resource Manager (SRM). gLite treats resources as Compute Elements (CE) or Storage Elements (SEs) depending on whether they are computational or data resources respectively. Jobs are generally non-interactive and batch oriented. The gLite Workload Management System (WMS) handles job scheduling and resource allocation and uses Condor-G for job dispatch and management. The WMS accepts job requests and stores them in its Task Queue. A Matchmaker sub-component matches job requests against resource information stored in an Information Super Market (ISM) sub-component, using the Condor ClassAds mechanism. The WMS uses both eager scheduling (jobs are 'pushed' to the resource) and lazy scheduling (resource 'pulls' or requests for jobs). Data required by a job scheduled at a CE is replicated to the nearest SE.

gLite works within a standardised Grid environment running EGEE middleware and has a standardised client configuration that requires external services such as R-GMA (Relational Grid Monitoring Architecture) Information System. gLite is installed on a dedicated machine and accepts job requests from local and remote clients. Thus, it is a centralised resource brokering system and therefore, differs considerably from the other resource brokers which are primarily user-directed, client-focused resource brokering mechanisms.

gLite automatically schedules replication of the required data for a job to the closest Storage Element to the Compute Element where the job has been scheduled. But, the locations of the data are not taken into account during the selection of Compute Element itself. That is, gLite does not perform any optimisation for reducing the amount of data to be transferred for an execution. gLite interfaces with an Accounting module that enables it to keep track of usage and charge users. However, it does not provide any economy-based scheduling of Grid applications.

## 4.5 Comparison

Table 4.1 compares the Gridbus broker and the related work discussed previously against characteristics derived from the challenges listed at the beginning of this chapter. While it may seem unfair to compare the other brokers against requirements that they were not designed for, this comparison is only a discussion of how the design of the Gridbus broker is different and not a measure of the applicability of the brokers to any situation.

From the table, it can be seen that the design of the Gridbus broker was motivated by different considerations than that of the other brokers. The focus of the Gridbus broker has been on scheduling and executing distributed data-intensive applications on potentially heterogeneous Grid resources. This is in contrast to Condor-G and gLite that are primarily job management systems, or Nimrod/G, that focuses on computationally intensive parameter sweep applications. APST schedules jobs so as to reuse data that has already been transferred but the initial location of the data is the client machine or the machine on which the broker is executing. Also, the Gridbus broker has been designed to enable economy-based strategies for Grid scheduling, going beyond the resource pricing provided by Nimrod-G, by supporting services such as market directories and resource accounting. The Gridbus broker is a single user system, each application execution requires a different instantiation of the broker. This is different to systems such as gLite which is a centralised resource broker that handles multiple users.

*Table 1: Comparison between the broker and related work*

| Characteristics | Nimrod/G | APST | Condor-G | gLite | Gridbus |
|---|---|---|---|---|---|
| **I Service Heterogeneity** | | | | | |
| 1. Support for different low-level computational middleware | Through Globus, Legion and SSH | Through Globus and SSH | Through Globus and Condor | Only EGEE | Through Globus and SSH as well as Alchemi |
| 2. Support for different Data Grid middleware | None | FTP, GridFTP | Through Stork | Only EGEE | FTP, GridFTP,SRB |
| 3. Equality of different service types | No | No | No | No | Yes |
| **II Support for Application Models** | | | | | |
| 1. Basic application model | Parameter Sweep | Parameter Sweep | Single Job | Single Job | Independent Tasks ( Jobs and Parameter Sweep ) |
| 2. Support for workflows | External | External | External | External | Internal |
| 3. Access to internal entities | Through database | Restricted API | Restricted API | Client API | Fully Programmable |
| **III Realisation of User Objectives** | | | | | |
| 1. Scheduling based on location of data | No | Yes | No | No | Yes |
| 2. Third-party data transfers. | Yes | Yes | Through Stork | Data Replication | Yes |
| 3. Late binding of data locations to jobs | No | Yes | Through Stork and Kangaroo | No | Yes |
| 4. Access to dynamic network information | Yes | Yes | No | No | Yes |
| 5. Resource pricing and cost-based scheduling | Yes | No | No | No | Yes |
| 6. Managing multiple credentials across VOs | No | No | No | No | Yes |
| **IV Infrastructure** | | | | | |
| 1. Checkpointing of jobs | Yes | No | Yes | Through Condor-G | No |
| 2. Execute-once semantics | No | No | Yes | -do- | Yes |
| 3. Local persistent store | Yes | Yes | Yes | Yes | Yes |
| 4. Dependencies on remote services | Yes | No | No | Yes | No |

One of the design principles that differentiate the Gridbus broker from the other resource brokers is the support for different Grid middleware. Except APST, the others work with only Globus services, or in the case of gLite, with resources running only EGEE middleware. This decoupling in the Gridbus broker has been achieved by limiting all middleware dependencies to the Execution layer and by not assuming the presence of specific services or libraries on the remote resources. The benefit of this loose coupling of the broker to low-level Grid middleware is that it can utilize a greater number and range of resources. The object-oriented nature of the broker also makes it easy to support any low-level Grid middleware, if required. For example, for compute Grid middleware all that is required is to extend the ComputeServer and the JobWrapper classes. However, this approach has its disadvantages as well. As mentioned before, Condor-G is able to provide stronger fault tolerance semantics due to its close integration with Globus and because it is able to install a virtualisation layer that periodically saves the state of jobs on the resources. Such a feature would require the broker to assume the availability of certain libraries on the resources.

Another distinctive design feature of the Gridbus broker is the equality of all types of services, whether they are compute, data or information services. That is, all the Grid services are treated as first-class citizens. This enables the broker to achieve different kinds of strategies such as those which give more prominence to data rather than computational requirements. The other brokers, with the exception of APST, focus on the computational aspect of the jobs. While these handle data in different ways - for example, Condor-G presents the data requirements of an application to Stork to handle while gLite simply replicates it on demand - they do not generally have strategies to choose a specific data repository at runtime based on current network conditions. The Gridbus broker has been designed to provision for such requirements.

# 5 Performance Evaluation

The Gridbus broker has been evaluated in the context of application deployment in several case studies published previously ([13],[32],[34],[53],[54]). In this section, however, we evaluate the performance of the internal elements of the broker alone, both on the client machine and in its interactions with remote middleware. From the previous discussion, it can be seen that the broker is designed to be a general purpose resource broker for executing large, data-intensive, coarse-grained distributed applications over global Grids. That is, its design was determined by generic requirements of distributed Grid applications and was not tied down to a specific application or remote Grid middleware. The first set of experiments examines the implications of this design on the performance of the broker on the machine on which it is running. However, the broker's performance is not only determined by its architecture and design but also by its interactions with the remote middleware. The second set of experiments evaluates the various overheads incurred due to meta-scheduling and remote-execution on the jobs.

In these experiments, we used a 'synthetic' batch job that computes a mathematical function and enables us to vary the job length and data files used for simulating long-running and data-intensive jobs respectively.

## 5.1 Performance of the Broker on its Host

The workload of the broker is determined by its Job and the Service objects. Therefore, the experiments that follow are divided into two parts: one with increasing number of jobs, and the other with increasing number of compute resources, or ComputeServer objects. The metrics measured were the memory occupied by the broker (HeapSize) and the number of threads produced by the broker. The broker was executed as a command-line program on a machine with Intel Xeon CPU and 2 GB of RAM running Redhat Linux 8.0 with version 1.4.3 of the Java Virtual Machine (JVM) installed. The remote computing resources used in this experiment are given in Table 2 and contain 3 clusters that are running job management systems such as PBS and SGE in conjunction with Globus.

Figure 13 shows the graphs of the memory occupied by the broker and the number of threads spawned by the broker averaged over the period of a single execution. In Figure 13(a), the number of jobs was increased from 20 to 2000 and a single compute resource was used throughout while in Figure 13(b), the number of compute resources was increased from 1 to 4 and the number of jobs was kept constant at 1000.

*Table 2: The set of Grid resources used for the experiment.*

| **Organization** | **Node details (Hostname, Architecture)** | **Grid Middleware** |
|---|---|---|
| University of Melbourne, Australia | `belle.cs.mu.oz.au` `IBM eServer, 4 CPU, 2GB RAM, 70 GB HD, RH Linux 8.0` | Globus 2.4 |
| University of Melbourne, Australia | `manjra.csse.unimelb.edu.au,` `Linux Cluster, 13 nodes` | Globus 4.0 + PBS |
| Victorian Partnership for Advanced Computing, Melbourne | `ng1.vpac.org,` `Linux Cluster, 16 nodes,` | Globus 2.4 + PBS |
| Universidad Complutense Madrid, Spain | `aquila.dacya.ucm.es,` `Linux Cluster` | Globus 4.0 + SGE |
| Universidad Complutense Madrid, Spain | `draco.dacya.ucm.es` | Globus 4.0 |

It can be seen that the average number of threads remains in a narrow range of 25-27 for increasing number of jobs while it increases steadily, though slowly, for increasing number of compute resources. The only threads within the broker that are constantly running throughout an execution are the ServiceMonitor, the Scheduler and the JobMonitor. These perform tasks that involve interaction with remote resources such as probing a compute resource, dispatching a batch of jobs and probing remote jobs by spawning time-limited transient threads. This is so because remote interactions are prone to failures and may cause bottlenecks within the broker if carried out in a single sequence. When the number of compute resources increases, the number of transient threads probing them increases as well.

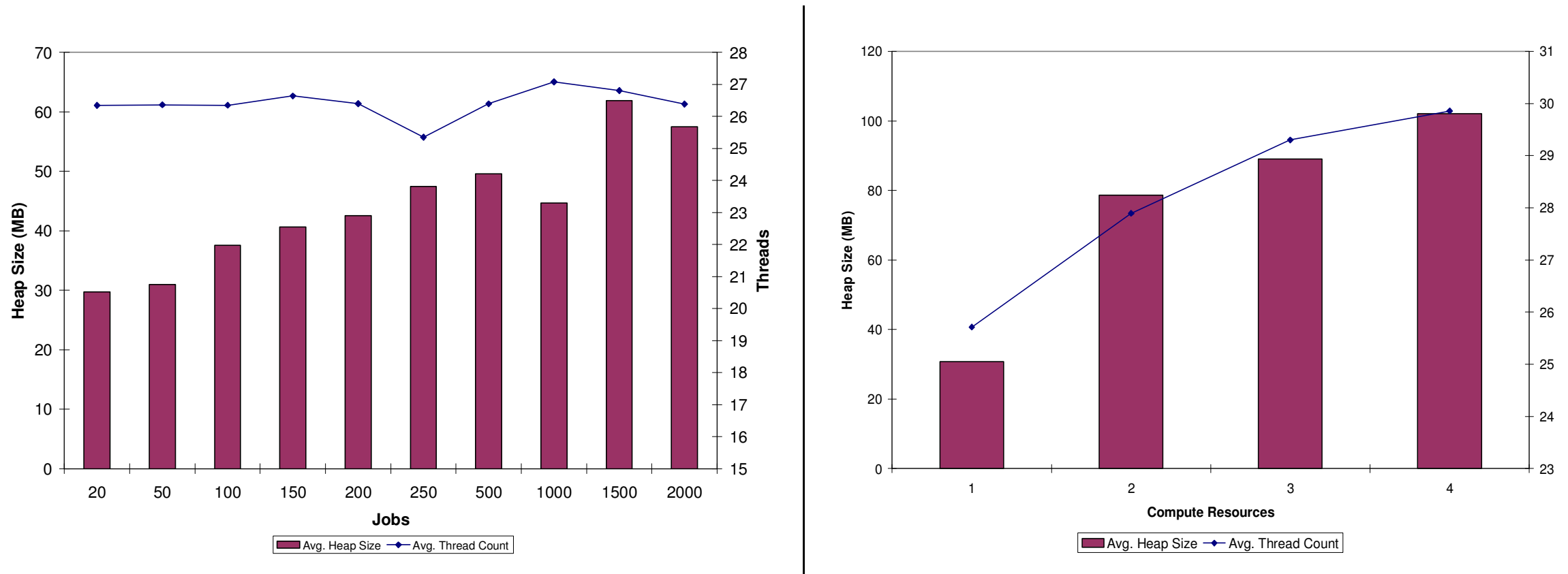

**(a)** **(b)**

*Figure 13: Avg. Heap Size and Avg. Thread Count for the broker for (a) increasing number of jobs and (b) increasing number of compute resources.*

The average memory occupied by the broker is shown to increase for both increasing number of jobs and compute resources. In case of jobs, the memory occupied increases from 29 MB to 57 MB when the number of jobs increases from 20 to 2000. Therefore, there is only a 2 times increase in memory size for a 100 times increase in workload. This is because the Scheduler loads only a subset of jobs (which we term as the “active set”) from the database into the memory at a time and this reduces the memory occupied by Job objects waiting to be executed. The size of the active set is configurable by the user and also depends on the memory available for the broker. In case of computing resources, the corresponding objects (ComputeServices) reside in the memory throughout an execution. Therefore, the memory usage increases with increasing number of services.

## 5.2 Interaction with Remote Middleware

Two experiments were performed; one for measuring various parameters such as job-waiting time, submission and monitoring time, actual execution time, and termination time under varying job-profiles, and the other measured the same under different middleware. Specifically, the following metrics were measured:

1. **Time for submission:** This is measured as the time span between the instant when the job was 'mapped' to a grid resource by the broker scheduler and the instant when the job handle was retrieved from the remote middleware system. This measure provides the time taken by the dispatcher to submit a mapped job to its chosen compute resource.
2. **Total querying time:** This is the total time spent for a job in querying its status on the remote node. It is measured as the sum of the times taken for each query at periodic intervals.
3. **Time for termination:** This is the time spent for terminating and cleaning up after a job is indicated as completed by the remote middleware. This, therefore, includes the time to retrieve the output files from the resource and the time to delete the directories created by the job on the remote resource.
4. **Job wallclock time:** This is measured as the difference between the time the job was submitted to the remote resource and the time the job was completed.

Each of the metrics mentioned is reported as a 'per job' value calculated as an average over a set of 50 jobs. The first experiment used one of the Grid resources from Table 2 (`ng1.vpac.org`) running Globus 2.4. The broker was executed as a non-interactive command-line program on a notebook computer (running Microsoft Windows XP, with an Intel Pentium M 2Ghz processor and 2GB RAM). The application binaries were installed on the remote middleware, and hence the times shown do not include the time required for actually copying the application binary files. The tests were run against different job profiles shown in Table 3.

*Table 3:Profile of jobs used in experiments*

| Job-profile | Job length (minutes) (approx.) | Data requirements (I/O) (MB) |
|---|---|---|
| Simple | 0.5 | < 1 |
| Data-intensive | 5 | 100 (approx.) |
| Compute-intensive | 10 | < 1 |

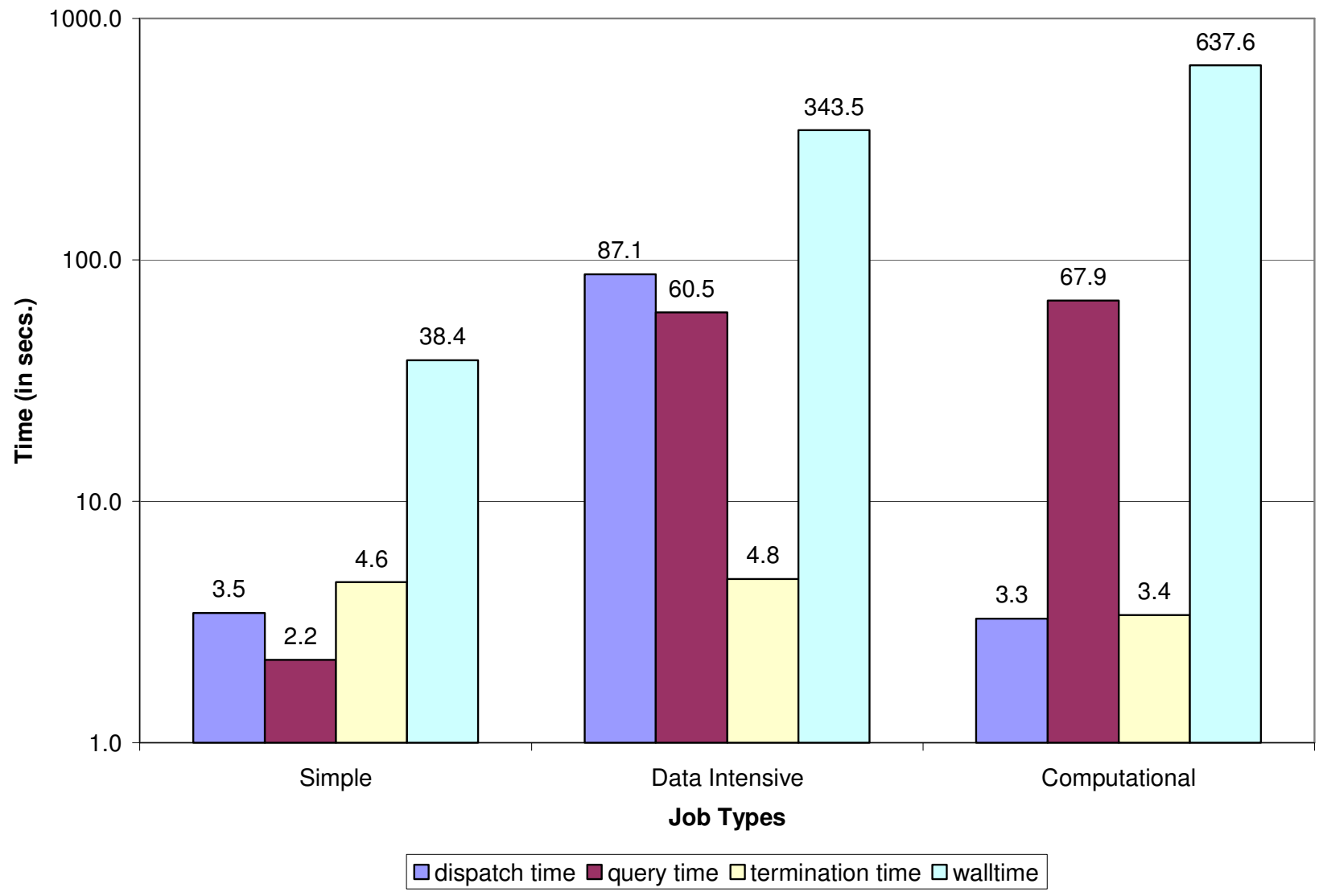

*Figure 14: A comparison of remote interaction overheads for different job types (GT2)***.**

Figure 14 shows the distribution of time taken by simple, data-intensive and compute-intensive jobs in various stages of a job execution on a Globus 2.4 resource. The time to dispatch a job to the remote resource is similar for both simple and computationally-intensive jobs but is higher for data-intensive jobs as it includes the time to stage in the input data file to the remote node. The jobs were queried every 12 seconds and therefore, the aggregate time spent by the broker to query a job naturally increases with the length of the job. This measure does not affect the wallclock time as the querying occurs on the broker's end. The termination time is almost equal for all the jobs as only the standard output and standard error files for the job are copied in all the three cases.

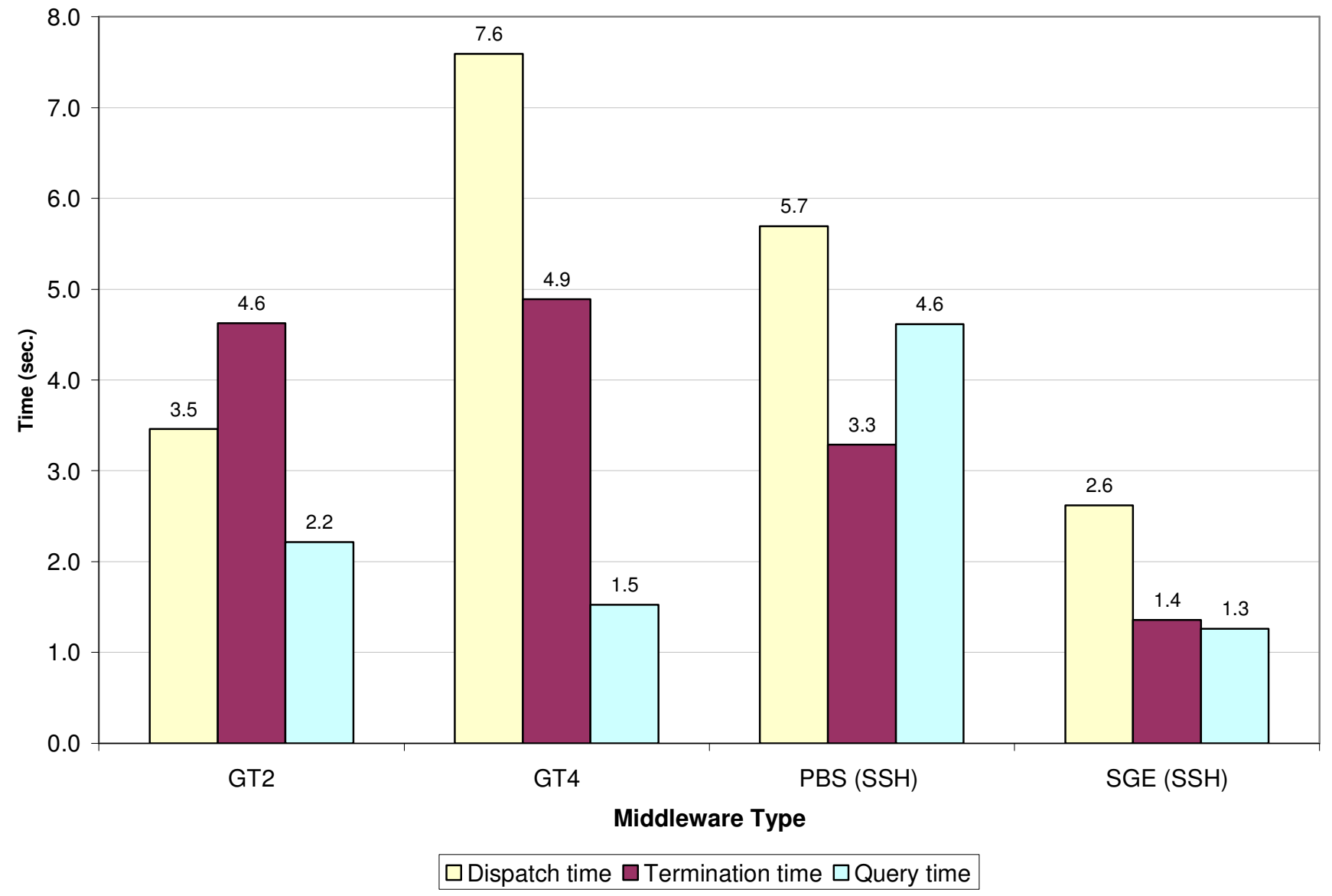

*Figure 15: Comparison of dispatch, termination and query times for different middleware on a single machine.*

The last experiment was performed by installing four low-level grid middleware - Globus 2 (GT2), Globus 4 (GT4), PBS and SGE - on a single machine which in this case was a cluster in our department (`manjra.csse.unimelb.edu.au`). The broker was started on another server machine within our department (`belle.csse.unimelb.edu.au`). All the four middleware were tested with a set of 50 jobs following the simple profile in Table 3. PBS and SGE were tested by direct invocation of commands through SSH. Figure 15 displays the results of this experiment.

It can be seen here that the highest dispatch time is observed for GT4, which could be due to the effect of web-services based messaging, resulting in extra delays for serialising and deserialising XML messages. GT2 incurs a higher overhead that SGE, which is probably due to the additional burden of security and encryption offered by the GSI libraries. The PBS and SGE dispatchers in the broker were expected to perform on par, however, we see that PBS has incurred higher overheads in all cases.

# 6 Lessons Learnt from our Experience

The development of the Gridbus broker with focus on data-intensive applications was initiated in early 2003; since then, it has undergone several version changes and feature additions. While the three layer architecture has remained unchanged from the very beginning of the broker, the design has been updated several times to include new features and to increase performance. We summarize the lessons that we have learnt in this interval in the following points:

1. **Dealing with Heterogeneity.** As mentioned in Section 3.3.1, the broker does not follow a specific interaction model with remote middleware or assume the availability of certain libraries on the resources. The presence of Grid middleware also presented with another layer of heterogeneity on top of the one caused by the presence of multiple operating systems. For example, Globus, which is the most popular Grid middleware in terms of deployment, has progressed from version 2.x through to 4.x over the space of 5 years. Each of these version changes have involved paradigm shifts- from RPC (Remote Procedure Calls) to OGSI (Open Grid Service Infrastructure) to WSRF (Web Services Resource Framework)-and are therefore, mostly incompatible with each other. Large Grid infrastructures such as UK eScience project, TeraGrid in the US and Australia's own APAC Grid have struggled to keep up with these changes and to port Grid applications from one version to another. In many cases, for example, in the EGEE project in Europe and the LHC Grid project, Globus 2.x is still used at the time of writing due to the massive investment in applications, personnel and training around the existing infrastructure. The Gridbus broker was initiated at a time when OGSI had only been just introduced. Therefore, its architecture and the design evolved out of the need to easily port between different versions of Grid middleware and was later generalized to cover all different types of middleware.

2. **Thread control.** The first version of the broker created a thread for every active job. However, as we realized soon, this was not scalable and also caused problems with thread synchronization. Also, when used within Web portals, the broker would not exit cleanly sometimes as some of the threads would remain suspended. Eventually, we re-designed the core to have only three main threads which fire off short-lived threads when interacting with remote resources. As mentioned earlier, the transient threads are also provided with time-outs to ensure that they will exit and be removed from the virtual machine.
3. **Handling multiple broker instances.** Up to version 2.2 of the broker, a properties file was provided to set up certain user-configurable parameters (e.g., location of credentials, polling time). Reference to this file (as a Java Properties object) was made static so that any component within the broker was able to access this file during runtime. However, this caused severe problems when 2 or more brokers were initiated within a Web portal container as they shared the reference to the same Properties object within the same Virtual Machine. This was remedied by providing persistence and a unique ID for each instance of the broker so that the properties can be referenced by the ID of the instance.
4. **Managing distributed development.** The number of developers working on the broker has steadily increased over the course of its development. While most of the contributions to the code have come from internal developers (i.e., associated with the GRIDS Lab), there have been a few external contributions. Also, some of the developers are no longer associated with the projects that have contributed to the broker. To manage this scenario, we set up a CVS repository to begin with and enforced guidelines for checking code into it. However, within the broker itself, there were different styles of coding, usage of non-standard variable names and "hacks" to enable certain features. Code cleanup is therefore, performed periodically by the available developers by examining the code together.

It has to be noted that issues 2 and 3 in the list above arose when the broker was applied to the portal environment. Prior to this, it was only used in command line environments where there was only one instance of the broker running inside a single VM. This is an example of how adapting to different user environments involves solving different challenges.

## 7 Summary and Conclusion

In this paper we have presented the architecture and design of Gridbus Broker that is motivated by the challenges and requirements of presenting an abstract interface for users and applications to distributed and heterogeneous Grid resources. We have shown how the presented design meets these requirements and how these differentiate it from the designs of similar Grid resource brokers. We have also presented two experiments that measured the effect of the design choices both on the performance of the broker on its host machine and in its interactions with remote resources. Lastly, we presented the lessons learnt from our development experience.

To conclude, it can be stated that the design of the Gridbus broker has been successful in meeting its requirements. It is also flexible enough to meet future challenges. For example, the broker was recently modified to provide an environment for describing and executing workflows based on work previously performed in the GRIDS Lab [55]. This was done without modifying the existing entities. This exercise will be subject of a future publication. The Gridbus broker therefore, provides an environment which users and developers can exploit to derive the maximum out of the available Grid services without having to worry about dealing with the challenges of heterogeneity and dynamic variations in availability.

## Acknowledgements

We would like to thank Prof. Christoph Reich (Furtwangen University, Germany) for his help with the profiling of the broker on the local machine. We would also like to thank VPAC and University Complutense Madrid for allowing us to use their resources for our experiments. We are grateful to David Abramson (Monash University, Australia) and Henri Casanova (University of Hawai`i, USA) for their comments regarding the comparison between the Gridbus broker and related systems. We would like to acknowledge all the developers who have contributed to the Gridbus broker over the years and the users who have provided suggestions that have led to many enhancements. This work is partially supported by a Discovery Project grant from the Australian Research Council.